\documentclass[a4paper,oneside,11pt]{article}

\usepackage[cp1250]{inputenc}

\newtheorem{dfn}{Definition}[section]
\newtheorem{tw}[dfn]{Theorem}
\newtheorem{prop}[dfn]{Proposition}
\newtheorem{rem}[dfn]{Remark}

\newtheorem{ex}[dfn]{Example}
\newtheorem{lem}[dfn]{Lemma}
\newtheorem{ass}[dfn]{Assumption}

\usepackage{anysize}
\usepackage{verbatim}
\usepackage{array}
\usepackage{enumerate}
\usepackage{amssymb} 
\usepackage{amsmath}
\usepackage{fancyhdr}
\usepackage{euscript}
\usepackage{latexsym}

\author{Micha\l \ Barski \\ \small  Faculty of Mathematics and Computer Science, University of Leipzig, Germany\\
\small Faculty of Mathematics, Cardinal Stefan Wyszy\'nski University in Warsaw, Poland
\\ \small{\it Michal.Barski@math.uni-leipzig.de}}

\title{\bf Asymptotic pricing in large financial markets}

\begin{document}

\maketitle

\begin{abstract}
The problem of hedging and pricing sequences of contingent claims in
large financial markets is studied. Connection between asymptotic
arbitrage and behavior of the $\alpha$~-~quantile price is shown.
The large Black-Scholes model is carefully examined.
\end{abstract}

\renewcommand{\abstractname}{Abstract}

\begin{quote}


\textbf{GEL Classification Numbers}: G11, G13
\end{quote}

\section{Introduction}

\label{1} \noindent \phantom{i}\qquad A large financial market is a
sequence of small arbitrage-free markets. Absence of arbitrage
opportunity on each element of the sequence does not guarantee that
there is no arbitrage ''in the limit''. Different concepts of
asymptotic arbitrage were introduced in~ \cite{KabKra1}
and~\cite{KabKra2} and their connections with some properties of
measure families: contiguity and asymptotic separation were shown.
For other similar results in this field see
also~\cite{Klein},~\cite{KleScha}. For other notions as asymptotic
free lunch and its relation with existence
of a martingale measure for the whole market see~\cite{Klein}, ~\cite{Ras}.%
\newline
Another problem arises as a natural consequence of asymptotic
arbitrage theory: how one can calculate the price of a contingent
claim and what is the connection between the price and the
no-arbitrage property of the market. We formulate the problem of
pricing not for a single random variable but for the sequence of
random variables instead. Motivation for such problem stating is
presented in section \ref{3}. For such a sequence we define
different types of sequences of hedging strategies. The first of
them hedges each element of the sequence, thus it carries no risk at
all. Basing on that property we define a strong price, which is
strictly related to the price known from the classical theory of
financial markets. The other type hedges the sequence with some risk
which does not exceed a fixed level in infinity. For this case we
introduce the $\alpha $-quantile price. In particular, the risk can
vanish in infinity indicating the $1$-quantile price which is called
a weak price. These definitions are presented in section
\ref{3}.\\
\phantom{i}\qquad In section \ref{4} we provide characterization
theorems for the prices mentioned above for general large
financial markets. This general description uses the no-arbitrage
property of each small market only. The question arises how the
prices are related to each other, in particular the strong and the
weak one, under different types of asymptotic arbitrage. Example
\ref{przyklad} shows that asymptotic arbitrage actually does
affect this relation. We study this problem and show a relevant
theorem for the sequence of complete markets. Analogous theorem
for incomplete markets remains an open problem.\\
\phantom{i}\qquad A significant part of the paper is section \ref{5}
devoted to the large Black-Scholes market with constant
coefficients. In these particular settings we improved previous
results and established more precise characterization theorems which
includes widely used derivatives such as call and put options. In
this section we also provide an alternative proof of the theorem
describing different kinds of asymptotic arbitrage which comes
from~\cite{KabKra2}. The method of proving is less general then
in~\cite{KabKra2}, but using Neyman-Pearson lemma provides more
indirect insight into the construction of relevant sets. Moreover,
similar methods based on non-randomized tests are
successfully used in other proofs in this section.\\
\phantom{i}\qquad The paper is organized as follows. In section
\ref{2} we present definitions of some properties of measure
families and known facts about asymptotic arbitrage. For a more
comprehensive exposition see~\cite {JacShi1} for the statistical
part and~\cite{KabKra1},~\cite{KabKra2},~\cite
{Klein},~\cite{KleScha} for the financial part. In section \ref{3}
we formulate precisely the problem of pricing. Section \ref{4}
provides characterization theorems which are used and generalized
in section \ref{5} describing the large Black-Scholes model.\\
\noindent \phantom{i}\qquad In general, the main idea in the $\alpha $%
-quantile price characterization theorem has its origin in the
paper on quantile hedging~\cite{FolLeu} . Thus the results
presented here can be treated as an extension or further
development in this field.

\section{Basic definitions and results}

\label{2} \noindent \phantom{i}\qquad By a large financial market we
mean a sequence of small markets. Let $(\Omega ^{n},\mathcal{F}^{n},(\mathcal{F}%
_{t}^{n}),P^{n})$, where $t\in [0,T^{n}]$ or $t\in
\{0,1,...,T^{n}\}$ be a sequence of filtered probability spaces
and $(S_{n}^{i}(t)),\ i=1,2,...,d_{n}$ a sequence of
semimartingales describing evolution of $d_{n} $ stock prices. A
large financial market will be called \textbf{stationary} if
$S_{n+1}^{i}(t)=S_{n}^{i}$ for $i=1,2,...,d_{n}$. This means that
each subsequent small market contains the previous one. To shorten
notation
assume that all the markets have the same time horizon, i.e. $T^{n}=T$ for $%
n=1,2,...$.\\
As a trading strategy on the $n$-th small market we admit a pair $%
(x_{n},\varphi _{n})$, where $x_{n}\geq 0$ and $\varphi _{n}$ is an $\mathbb{%
R}^{d_{n}}$ valued predictable process integrable with respect to
$(S_{n}(t)) $. The value of  $x_{n}$ is an initial endowment and
$\varphi _{n}^{i}(t)$ is a number of units of the $i$-th stock held
in the portfolio at time $t$. The wealth process corresponding to
the strategy $(x_{n},\varphi _{n})$ defined as $V_{t}^{x_{n},\varphi
_{n}}=\sum_{i=1}^{d_{n}}\varphi _{n}^{i}(t)S_{n}^{i}(t)$ is assumed
to satisfy a self-financing condition, that is:
\begin{gather*}
V_{t}^{x_{n},\varphi _{n}}=x_{n}+\int_{0}^{t}\varphi _{n}(t)dS_{n}(t)\ \text{%
for continuous time models} \\[2ex]
V_{t}^{x_{n},\varphi
_{n}}=x_{n}+\sum_{s=1}^{t}\sum_{i=1}^{d_{n}}\varphi
_{n}^{i}(s)(S_{n}^{i}(s)-S_{n}^{i}(s-1))\ \text{for discrete time
models}.
\end{gather*}

\begin{dfn}
A pair $(0,\varphi _{n})$ is an arbitrage strategy on the $n$-th
small market if $V_{t}^{0,\varphi _{n}}\geq 0$ a.s. for each $t$
and
\begin{equation*}
P^{n}(V_{T}^{0,\varphi _{n}}>0)>0.
\end{equation*}
\end{dfn}
\noindent
For the $n$-th small market we recall the definition of the set $\mathcal{Q}%
^{n}$ of all martingale measures.

\begin{dfn}
$Q\in \mathcal{Q}^{n}\Longleftrightarrow (S_{n}^{i}(t))$ is a
local martingale on $[0,T]$ with respect to $Q$ for
$i=1,2,...,d_{n}$
\end{dfn}

\begin{tw}
If $\mathcal{Q}^{n}\neq \emptyset $ then there is no arbitrage
strategy on the $n$-th small market.
\end{tw}
\noindent
The proof can be found in~\cite{JacShi1} for discrete
time and in~\cite {DelSch} for continuous time settings. It turns
out that the inverse statement remains true for the discrete case,
but is false for continuous time.\newline

\noindent Throughout all the paper we assume that:
\begin{gather*}
\mathcal{Q}^n\neq\emptyset \quad \text{for all} \quad n=1,2,... \
.
\end{gather*}

\noindent The fact that there is no arbitrage on each small market
does not guarantee that there is no asymptotic arbitrage
opportunity. For the large financial markets we have the following
concepts of asymptotic arbitrage which comes from ~\cite{KabKra2}.

\begin{dfn}
A sequence of strategies $(x_{n},\varphi _{n})$ realizes the
asymptotic arbitrage of the first kind (AA1) if:
\begin{align*}
& V_{t}^{x_{n},\varphi _{n}}\geq 0\ \text{for all}\ t\in \lbrack 0,T] \\
& \lim_{n}x_{n}=0, \\
& \lim_{n}P^{n}(V_{T}^{x_{n},\varphi _{n}}\geq 1)>0.
\end{align*}
\end{dfn}

\begin{dfn}
A sequence of strategies $(x_{n},\varphi _{n})$ realizes the
asymptotic arbitrage of the second kind (AA2) if:
\begin{align*}
& V_{t}^{x_{n},\varphi _{n}}\leq 1\ \text{for all}\ t\in \lbrack 0,T] \\
& \lim_{n}x_{n}>0, \\
& \lim_{n}P^{n}(V_{T}^{x_{n},\varphi _{n}}\geq \varepsilon )=0\ \text{for any%
}\ \varepsilon >0.
\end{align*}
\end{dfn}

\begin{dfn}
A sequence of strategies $(x_{n},\varphi _{n})$ realizes the
strong asymptotic arbitrage of the first kind (SAA1) if:
\begin{align*}
& V_{t}^{x_{n},\varphi _{n}}\geq 0\ \text{for all}\ t\in \lbrack 0,T] \\
& \lim_{n}x_{n}=0, \\
& \lim_{n}P^{n}(V_{T}^{x_{n},\varphi _{n}}\geq 1)=1.
\end{align*}
\end{dfn}

\begin{dfn}
A sequence of strategies $(x_{n},\varphi _{n})$ realizes the
strong asymptotic arbitrage of the second kind (SAA2) if:
\begin{align*}
& V_{t}^{x_{n},\varphi _{n}}\leq 1\ \text{for all}\ t\in \lbrack 0,T] \\
& \lim_{n}x_{n}=1, \\
& P^{n}(V_{T}^{x_{n},\varphi _{n}}\geq \varepsilon )=0\ \text{for
any}\ \varepsilon >0.
\end{align*}
\end{dfn}
\noindent We say that the large financial market does not admit the
asymptotic arbitrage of the first kind (second kind, strong
asymptotic arbitrage of the first kind, strong asymptotic arbitrage
of the second kind ) and denote this property by $NAA1$, ($NAA2$
,$NSAA1$, $NSAA2$) if for any sequence $(n_{k})$ there are no
trading strategies $(x_{n_{k}},\varphi _{n_{k}})$ realizing the
corresponding kind of asymptotic arbitrage. \newline For
characterization of the asymptotic arbitrage and for later purposes
we introduce some definitions from mathematical statistics.

\begin{dfn}
Let $(\Omega ^{n},\mathcal{F}^{n}),n=1,2,...$ be a sequence of
measurable spaces and $G_{n},H_{n}:\mathcal{F}^{n}\
\longrightarrow \ \mathbb{R}_{+}$ a sequence of set
functions.\newline
$1)$ $(G_{n})$ is \textbf{contiguous} with respect to $(H_{n})$ (notation: $%
(G_{n})\vartriangleleft (H_{n})$) if for every sequence $A_{n}\in \mathcal{F}%
^{n}$ we have
\begin{equation*}
H_{n}(A_{n})\longrightarrow 0\quad \Longrightarrow \quad
G_{n}(A_{n})\longrightarrow 0
\end{equation*}
$2)$ $(G_{n})$ is \textbf{asymptotically separable} from $(H_{n})$ (notation:%
$(G_{n})\vartriangle (H_{n})$) if there exists a sequence $A_{n}\in \mathcal{%
F}^{n}$ such that
\begin{equation*}
H_{n}(A_{n})\longrightarrow 0\quad \text{and}\quad
G_{n}(A_{n})\longrightarrow 1
\end{equation*}
\end{dfn}
\noindent
For the family $\mathcal{Q}^{n}$ we consider the
following set functions:
\begin{gather*}
\bar{\mathbf{Q}}^{n}(A)=\sup_{Q\in \mathcal{Q}^{n}}Q(A),\ A\in \mathcal{F}%
^{n}\quad \text{ - the upper envelope of}\ \mathcal{Q}^{n} \\
\underline{\mathbf{Q}}^{n}(A)=\inf_{Q\in \mathcal{Q}^{n}}Q(A),\
A\in \mathcal{F}^{n}\quad \text{ - the lower envelope of}\
\mathcal{Q}^{n}.
\end{gather*}
\noindent The following result provides characterization of
asymptotic arbitrage in terms of sequences of sets. For the proofs
see ~\cite{KabKra1},~ \cite{KabKra2},~\cite{KleScha}.

\begin{tw}
\label{tw o arbitrazu asymtotycznym} The following conditions hold

\begin{enumerate}
\item  (NAA1) \ iff \ $(P^{n})\vartriangleleft
(\bar{\mathbf{Q}}^{n})$

\item  (NAA2) \ iff \
$(\underline{\mathbf{Q}}^{n})\vartriangleleft (P^{n})$

\item  (SAA1) \ iff \ (SAA2) \ iff \ $(P^{n})\vartriangle (\bar{\mathbf{Q}}%
^{n})$ \ iff \ $(\underline{\mathbf{Q}}^{n})\vartriangle (P^{n})$.
\end{enumerate}
\end{tw}

Below we present a standard tool from mathematical statistics for
searching
optimal tests. It is useful to solve the following problem. Let $Q_{1}$ and $%
Q_{2}$ be two probability measures with density
$\frac{dQ_{1}}{dQ_{2}}$ on a
measurable space $(\Omega ,\mathcal{F})$. We are interested in finding set $%
\tilde{A}$ , which is a solution of the problem
\begin{equation*}
A\in \mathcal{F}:
\begin{cases}
Q_{1}(A)\longrightarrow \max  \\
Q_{2}(A)\leq \gamma
\end{cases}
\end{equation*}
with $\gamma \in \lbrack 0,1]$. Then the explicit solution is
given by the following lemma.

\begin{lem}
(\textbf{Neyman-Pearson})\label{lem NP}\newline
If there exists constant $\beta $ such that $Q_{2}\{\frac{dQ_{1}}{dQ_{2}}%
\geq \beta \}=\gamma $ then $Q_{1}\{\frac{dQ_{1}}{dQ_{2}}\geq
\beta \}\geq Q_{1}(B)$ for any set $B$ satisfying $Q_{2}(B)\leq
\gamma $.
\end{lem}

We recall also the pricing theorem, which has its origin in the
theorem on optional decomposition of the supermartingales. For more
details see~\cite{Kra} and for later
extensions~\cite{FolKab},~\cite{FolKra}.

\begin{tw}[Price characterization]
\label{tw o cenie} Let $\mathcal{Q}$ be a set of martingale
measures for the semimartingale $(S_{t})$ describing evolution of
the stock prices. Let $H$
be a non negative contingent claim. Then there exists a trading strategy $(%
\tilde{x},\tilde{\varphi})$, where $\tilde{x}=\sup_{Q\in \mathcal{Q}}\mathbf{%
E}^{Q}[H]$ s.t.
\begin{equation*}
\tilde{x}+\int_{0}^{t}\tilde{\varphi}(s)dS(s)\geq \underset{Q\in \mathcal{Q}%
}{\emph{ess}\sup }\ \mathbf{E}^{Q}[H\mid \mathcal{F}_{t}].
\end{equation*}
The pair $(\tilde{x},\tilde{\varphi})$ is thus a hedging strategy and $%
\tilde{x}$ is the price of $H$.
\end{tw}

\section{Problem formulation}

\label{3}

\begin{dfn}
A contingent claim $\mathbb{H}$ on a large financial market is a
sequence of
random variables $H_{1},H_{2},...$ satisfying the following conditions%
\newline
$1)$ For each $n=1,2...$ \ $H_{n}:\Omega ^{n}\longrightarrow
\mathbb{R}^{+}$ is an $\mathcal{F}^{n}$ measurable, non negative
random variable. \newline $2)$ For each $n=1,2,...$ $\sup_{Q\in
\mathcal{Q}^{n}}\mathbf{E} ^{Q}[H_{n}]<\infty $ holds.
\end{dfn}
\noindent In classical market models we have always one random
variable which we want to price and hedge. The question arises for
justification of considering a sequence of random variables. We
present two motivations for this fact.

\begin{enumerate}[1)]
\item  Assume that we have one random variable $G$ which is
measurable with respect to the $\sigma$-field $\sigma
(\mathcal{F}^{1},\mathcal{F}^{2},...)$. Then $H_{n}$ can
be defined as projections of $G$ on the spaces $(\Omega ^{n},\mathcal{F}%
^{n},P^{n})$, i.e. $H_{n}=\mathbf{E}^{P^{n}}[G\mid
~\mathcal{F}^{n}]$. Thus, we want to price a derivative which
depends on infinitely many assets but taking into account
information which is provided by the few coming first.

\item  Let $G$ be a random variable which depends on the price of
the first asset (or some first assets as well) only. Then we can
define $H_{n}=G$ for each $n$ and consider opportunity arising from
the fact that the number of assets which can be traded is
increasing. We examine how the increasing number of investments
possibilities affects the price of $G$.
\end{enumerate}

\noindent Below we present two concepts of asymptotic hedging and
prices definitions of $\mathbb{H}$.

\begin{dfn}
\label{def_mocej_ceny} A sequence $(x_{n},\varphi _{n})_{n}$ is a \textbf{%
sequence of hedging strategies} if
\begin{equation*}
V_{T}^{x_{n},\varphi _{n}}\geq H_{n}\quad \forall \ n=1,2,....
\end{equation*}
Such class of sequences we denote by $\mathcal{H}^{1}$. A
\textbf{strong price} of $\mathbb{H}$ is defined as
\begin{equation*}
v(\mathbb{H})=\inf_{(x_{n},\varphi _{n})\in \mathcal{H}^{1}}\ \underset{%
n\rightarrow \infty }{\underline{\lim }}x_{n}.
\end{equation*}
\end{dfn}

\noindent Throughout the whole paper we assume that $\alpha$ is
any number from the interval $[0,1]$.

\begin{dfn}
\label{def_slabej_ceny} A sequence $(x_{n},\varphi _{n})_{n}$ is a \textbf{%
sequence of $\alpha $-hedging strategies} if
\begin{equation*}
\underset{n\rightarrow \infty }{\underline{\lim
}}P^{n}(V_{T}^{x_{n},\varphi _{n}}\geq H_{n})\geq \alpha .
\end{equation*}
Such class of sequences we denote by $\mathcal{H}_{\alpha }$. An \textbf{$%
\mathbf{\alpha }$-quantile price} of $\mathbb{H}$ is defined as
\begin{equation*}
v_{\alpha }(\mathbb{H})=\inf_{(x_{n},\varphi _{n})\in
\mathcal{H}_{\alpha }}\ \underset{n\rightarrow \infty
}{\underline{\lim }}x_{n}.
\end{equation*}
A \textbf{weak price} of $\mathbb{H}$ is the {$1$-quantile price},
i.e.
\begin{equation*}
\tilde{v}(\mathbb{H}):=v_{1}(\mathbb{H}).
\end{equation*}
\end{dfn}

\noindent As follows from the definition above, we consider
sequences of strategies which do not allow to exceed a fixed level
of risk when $n$ tends to infinity. If $\alpha=1$, then the risk
vanishes in infinity. This particular case is distinguished to
compare with classical concept of pricing suggested by Definition
\ref{def_mocej_ceny}, where there is no risk for any $n=1,2,...
$.\newline \noindent At this stage it is clear that
$v_{\alpha}(\mathbb{H})\leq
v_{\beta}(\mathbb{H})\leq\tilde{v}(\mathbb{H})\leq v(\mathbb{H})$ for $%
\alpha<\beta$ since the following inclusions hold : $\mathcal{H}%
_{\alpha}\supseteq\mathcal{H}_{\beta}\supseteq \mathcal{H}_{1}\supseteq%
\mathcal{H}^{1}$. The main goal of the paper is to provide the
characterization of the prices and solve the problem of equality
between the strong and the weak price.

\section{Prices characterization}

\label{4} Using the price characterization Theorem \ref{tw o cenie}
on a classical market it is simple to show the following.

\begin{prop}
The strong price is given by
\begin{equation*}
v(\mathbb{H})=\underset{n}{\underline{\lim }}\ \sup_{Q\in \mathcal{Q}^{n}}%
\mathbf{E}^{Q}[H_{n}].
\end{equation*}
\end{prop}
{\bf Proof :} Let $g:=\underset{n}{\underline{\lim }}\ \sup_{Q\in
\mathcal{Q}^{n}}\mathbf{E}^{Q}[H_{n}]$. By Theorem \ref{tw o cenie},
for any $(x_{n},\varphi _{n})\in \mathcal{H}^{1}$ we get $x_{n}\geq
\sup_{Q\in
\mathcal{Q}^{n}}\mathbf{E}^{Q}[H_{n}]$ and thus $v(\mathbb{H})\geq g$.%
\newline
\noindent Taking $\tilde{x}_{n}:=\sup_{Q\in \mathcal{Q}^{n}}\mathbf{E}%
^{Q}[H_{n}]$, from Theorem \ref{tw o cenie} we know  that there
exists a
sequence of strategies $(\tilde{\varphi}_{n})$ s.t. $(\tilde{x}_{n},\tilde{%
\varphi}_{n})\in \mathcal{H}^{1}$ and thus $v(\mathbb{H})\leq g$. \hfill {$%
\square $}\newline
\newline
\noindent To characterize the weak price we introduce first some
definitions.

\begin{dfn}
(The class $\mathcal{A}_{\alpha }$)\newline A sequence of sets
$(A_{n})$ belongs to the class $\mathcal{A}_{\alpha }$ if
$A_{n}\in \mathcal{F}^{n}$ for $n=1,2,...$ and
$\underset{n\rightarrow \infty }{\underline{\lim
}}P^{n}(A_{n})\geq \alpha $. In particular $(A_{n})$
belongs to the class $\mathcal{A}_{1}$ if $P^{n}(A_{n})\underset{n}{%
\longrightarrow }1$.
\end{dfn}
\noindent
The following remarks state the correspondence between the class of $\alpha$%
-hedging sequences $\mathcal{H}_{\alpha}$ and the class of $\mathcal{A}%
_{\alpha}$ sets.

\begin{rem}
\label{rem 1}($\mathcal{A}_{\mathcal{H}_{\alpha }}\subseteq \mathcal{A}%
_{\alpha }$)\newline
Each element in $\mathcal{H}_{\alpha }$ indicates an element in $\mathcal{A}%
_{\alpha }$. Indeed, for $(x_{n},\varphi _{n})\in
\mathcal{H}_{\alpha }$ let us define $A_{n}^{x_{n},\varphi
_{n}}:=\{V_{T}^{x_{n},\varphi _{n}}\geq
H_{n}\}$. By definition of $\mathcal{H}_{\alpha }$ we obtain that $%
(A_{n}^{x_{n},\varphi _{n}})\in \mathcal{A}_{\alpha }$. Thus, if we
denote the sequences of sets above by
$\mathcal{A}_{\mathcal{H}_{\alpha }}$ the following inclusion holds
: $\mathcal{A}_{\mathcal{H}_{\alpha }}\subseteq \mathcal{A}$.
\end{rem}

\begin{rem}
\label{rem 2}($\mathcal{H}_{\mathcal{A}_{\alpha }}\subseteq \mathcal{H}%
_{\alpha }$)\newline For the sequence $(A_{n})\in
\mathcal{A}_{\alpha }$ let us consider a
sequence of strategies s.t. for a fixed number $n$ strategy $%
(x_{n}^{A},\varphi _{n}^{A})$ satisfies : $x_{n}^{A}=\sup_{Q\in \mathcal{Q}%
^{n}}\mathbf{E}^{Q}[H_{n}\mathbf{1}_{A_{n}}]$ and $\varphi _{n}^{A}$
hedges the contingent claim $H_{n}\mathbf{1}_{A_{n}}$(on a small
market with index $n$). It follows that $(x_{n}^{A},\varphi
_{n}^{A})\in \mathcal{H}_{\alpha }$ since $(A_{n})\in
\mathcal{A}_{\alpha }$. If we denote the sequences of strategies of
the form above by $\mathcal{H}_{\mathcal{A}_{\alpha }}$ the
following inclusion holds: $\mathcal{H}_{\mathcal{A}_{\alpha
}}\subseteq \mathcal{H}_{\alpha }$.
\end{rem}

\begin{tw}
\label{tw o cenie kwantylowej} The $\alpha $-quantile price is
given by
\begin{equation*}
v_{\alpha }(\mathbb{H})=\inf_{(A_{n})\in \mathcal{A}_{\alpha }}\ \underset{n%
}{\underline{\lim }}\ \sup_{Q\in \mathcal{Q}^{n}}\mathbf{E}^{Q}[H_{n}\mathbf{%
1}_{A_{n}}].
\end{equation*}
\end{tw}
{\bf Proof :} We show successively two inequalities: $(\geq)$\ \
and $(\leq )$.\\
($\geq $) \ Let us consider $(x_{n},\varphi _{n})\in
\mathcal{H}_{\alpha }$. Then using the notation of Remark \ref{rem
1} we have:
\begin{equation*}
x_{n}\geq \sup_{Q\in \mathcal{Q}^{n}}\mathbf{E}^{Q}[H_{n}\mathbf{1}%
_{A_{n}^{x_{n},\varphi _{n}}}]
\end{equation*}
and therefore
\begin{equation*}
\underset{n}{\underline{\lim }}\ x_{n}\geq
\underset{n}{\underline{\lim }}\
\sup_{Q\in \mathcal{Q}^{n}}\mathbf{E}^{Q}[H_{n}\mathbf{1}_{A_{n}^{x_{n},%
\varphi _{n}}}].
\end{equation*}
By the definition of the $\alpha $-quantile price and by Remark
\ref{rem 1} we obtain
\begin{align*}
v_{\alpha }(\mathbb{H})=\inf_{(x_{n},\varphi _{n})\in
\mathcal{H}_{\alpha }}\ \underset{n}{\underline{\lim }}\ x_{n}&
\geq \inf_{(x_{n},\varphi _{n})\in \mathcal{H}_{\alpha
}}\underset{n}{\underline{\lim }}\sup_{Q\in
\mathcal{Q}^{n}}\mathbf{E}^{Q}[H_{n}\mathbf{1}_{A_{n}^{x_{n},\varphi
_{n}}}]
\\[2ex]
& =\inf_{(A_{n})\in \mathcal{A}_{\mathcal{H}_{\alpha }}}\underset{n}{%
\underline{\lim }}\sup_{Q\in \mathcal{Q}^{n}}\mathbf{E}^{Q}[H_{n}\mathbf{1}%
_{A_{n}}] \\[2ex]
& \geq \inf_{(A_{n})\in \mathcal{A}_{\alpha }}\underset{n}{\underline{\lim }}%
\sup_{Q\in \mathcal{Q}^{n}}\mathbf{E}^{Q}[H_{n}\mathbf{1}_{A_{n}}]
\end{align*}
$(\leq )$ Consider an arbitrary element $(A_{n})\in
\mathcal{A}_{\alpha }$ and a corresponding strategy described in
Remark \ref{rem 2}. Following the notation of Remark \ref{rem 2} we
have
\begin{equation*}
\sup_{Q\in
\mathcal{Q}^{n}}\mathbf{E}^{Q}[H_{n}\mathbf{1}_{A_{n}}]=x_{n}^{A}
\end{equation*}
and therefore
\begin{equation*}
\underset{n}{\underline{\lim }}\sup_{Q\in \mathcal{Q}^{n}}\mathbf{E}%
^{Q}[H_{n}\mathbf{1}_{A_{n}}]=\underset{n}{\underline{\lim }}\
x_{n}^{A}
\end{equation*}
By Remark \ref{rem 2} we obtain
\begin{align*}
\inf_{(A_{n})\in \mathcal{A}_{\alpha }}\underset{n}{\underline{\lim }}%
\sup_{Q\in
\mathcal{Q}^{n}}\mathbf{E}^{Q}[H_{n}\mathbf{1}_{A_{n}}]&
=\inf_{(A_{n})\in \mathcal{A}_{\alpha }}\
\underset{n}{\underline{\lim }}\
x_{n}^{A} \\[2ex]
& =\inf_{(x_{n},\varphi _{n})\in \mathcal{H}_{\mathcal{A}_{\alpha
}}}\
\underset{n}{\underline{\lim }}\ x_{n} \\[2ex]
& \geq \inf_{(x_{n},\varphi _{n})\in \mathcal{H}_{\alpha }}\ \underset{n}{%
\underline{\lim }}\ x_{n} \\[2ex]
& =v_{\alpha }(\mathbb{H})
\end{align*}
$\hfill \square $ \newline \noindent We examine the problem of
asymptotic pricing studying the following example.

\begin{ex}
\label{przyklad} Let us consider the stationary large financial
market with the following settings:
\begin{equation*}
\Omega =[0,1],\quad S_{n}^{i}(1)=S_{n}^{i}(0)(1+\xi _{i}),\
i=1,2,...,n,\ n=1,2,...
\end{equation*}
where $(\xi _{i})$ is a sequence of random variables given by
\begin{equation*}
\xi _{i}=
\begin{cases}
-1\  & \text{on}\ [0,1-\frac{1}{2^{i}}]:=E_{i} \\
\frac{\delta (2^{i}-1)}{2^{i}-\delta (2^{i}-1)}\  & \text{on}\ (1-\frac{1}{%
2^{i}},1]:=F_{i},\quad \delta \in (0,1).
\end{cases}
\end{equation*}
Sigma fields are assumed to be generated by the sequence $(\xi _{i})$, i.e. $%
\mathcal{F}^{n}=\sigma (\xi _{1},\xi _{2},...,\xi _{n})$, and the
$n$-th objective probability measure $P^{n}$ is a restriction of
the Lebesgue's
measure $P$ on $[0,1]$ to the sigma-field $\mathcal{F}^{n}$, i.e. $%
P^{n}=P_{\mid \mathcal{F}^{n}}$. Each martingale measure $Q^{n}$ on the $n$%
-th market is described by the property : $\mathbf{E}^{Q^{n}}[\xi _{1}]=0,%
\mathbf{E}^{Q^{n}}[\xi _{2}]=0,...,\mathbf{E}^{Q^{n}}[\xi _{n}]=0$. Thus $%
Q^{n}$ is indicated by its values on the intervals
$E_{1},E_{2},...,E_{n}$ and one can check that
\begin{align*}
Q^{n}(E_{1})& =\delta \left( 1-\frac{1}{2}\right)  \\
Q^{n}(E_{2})& =\delta \left( 1-\frac{1}{2^{2}}\right)  \\
...&  \\
Q^{n}(E_{n})& =\delta \left( 1-\frac{1}{2^{n}}\right) .
\end{align*}
It follows from the above that we have constructed a sequence of
complete markets. \newline \noindent We shall find an {$\alpha
$}-quantile price of a trivial contingent claim $\mathbb{H}\equiv
1$.

\begin{prop}
In the model specified above we have:
\begin{equation*}
v_{\alpha }(1)=\delta \alpha .
\end{equation*}
\end{prop}
{\bf Proof :} We shall construct explicitly a sequence of sets $(\tilde{A}%
_{n})\in \mathcal{A}_{\alpha }$ satisfying:
\begin{equation*}
{\lim }\ Q^{n}(\tilde{A}_{n})=\inf_{(A_{n})\in \mathcal{A}_{\alpha }}{%
\underline{\lim }}\ Q^{n}(A_{n}).
\end{equation*}
\noindent Let:  $G_{1}:=E_{1}$, $G_{n}:=E_{n}\setminus E_{n-1}$
for n=2,3,....Then
\begin{gather*}
P^{n}(G_{n})=P^{n}(F_{n})=\frac{1}{2^{n}}\quad \mathit{and}
\\
\frac{\delta }{2^{n}}=Q^{n}(G_{n})<Q^{n}(F_{n})=1-\delta \Big(1-\frac{1}{%
2^{n}}\Big)
\end{gather*}
\noindent Consider a series expansion of $\alpha $:
\begin{equation*}
\alpha =\sum_{i=1}^{\infty }\frac{\gamma _{i}}{2^{i}},\ \ \ where\
\ \gamma _{i}\in \{0,1\}.
\end{equation*}
Define $\tilde{A}_{n}$ as follows
\begin{equation*}
\tilde{A}_{n}:=\bigcup_{i=1}^{n}\mathbf{1}_{\{\gamma
_{i}=1\}}G_{i}
\end{equation*}
and notice, that $P^{n}(\tilde{A}_{n})=\sum_{i=1}^{n}\gamma
_{i}P^{n}(G_{i})=\sum_{i=1}^{n}\frac{\gamma _{i}}{2^{i}}$ and therefore $%
\lim_{n\rightarrow \infty }P^{n}(\tilde{A}_{n})=\sum_{i=1}^{\infty }\frac{%
\gamma _{i}}{2^{i}}=\alpha $, so $(\tilde{A}_{n})\in
\mathcal{A}_{\alpha }$.
For any $(A_{n})\in \mathcal{A}_{\alpha }$ we have $\lim P^{n}(\tilde{A}%
_{n})\leq \underline{\lim }P^{n}(A_{n})$ and $\lim
Q^{n}(\tilde{A}_{n})\leq \underline{\lim }Q^{n}(A_{n})$.

\noindent Thus
\begin{align*}
v_{\alpha }(1)& =\inf_{(A_{n})\in \mathcal{A}_{\alpha }}\ \underset{n}{%
\underline{\lim }}\ \ \mathbf{E}^{Q^{n}}[\mathbf{1}_{A_{n}}]=\underset{n}{{%
\lim }}\ \ Q^{n}[\tilde{A}_{n}]=\underset{n}{{\lim }}\ \ \sum_{i=1}^{n}%
\mathbf{1}_{\{\gamma _{i}=1\}}{Q^{n}}(G_{i}) \\
& =\sum_{i=1}^{\infty }\ \gamma _{i}\frac{\delta }{2^{i}}=\delta
\sum_{i=1}^{\infty }\frac{\gamma _{i}}{2^{i}}=\delta \alpha .
\end{align*}
\hfill $\square $\newline
\noindent Notice that $\delta =\tilde{v}(1)<v(1)=\underset{n}{\underline{%
\lim }}\ \mathbf{E}^{Q^{n}}[1]=1$, so this example shows that
strict inequality between the strong and the weak price is
possible.\newline
\newline
\noindent Notice also that this model admits AA2 and does not
satisfy AA1.
Indeed, taking the sequence $(F_{n})$, we get: $P^{n}(F_{n})=\frac{1}{2^{n}}%
\longrightarrow 0$ and $Q^{n}(F_{n})=1-\delta \left( 1-\frac{1}{2^{n}}%
\right) \longrightarrow 1-\delta >0$ and thus there is AA2. Let
$(A_{n})$ be a sequence s.t. $Q^{n}(A_{n})\longrightarrow 0$. This
means that for any $l>0$, $Q^{n}(A_{n})<\frac{\delta }{2^{l}}$ holds
for all large $n$ and one can check, that this implies that
$A_{n}\subseteq (1-\frac{1}{2^{l}},1]$ for all large $n$. As a
consequence we obtain $\lim_{n}P^{n}(A_{n})<\frac{1}{2^{l}}$ and
letting $l$ to $\infty $ we get $\lim_{n}P^{n}(A_{n})=0$. This means
that NAA1 and  also NSAA1, NSAA2 hold.\newline
\newline
\noindent This example shows that NAA1, NSAA1, NSAA2 is
insufficient for the equality of the strong and the weak price.
\end{ex}

\noindent Theorem \ref{tw o cenie kwantylowej} yields immediately
two following conclusions.

\begin{rem}
If we require that $\tilde{v}(\mathbb{H})=v(\mathbb{H})$ even for
$\mathbb{H} $ of simple structure then the market must satisfy
$NAA2$. Indeed, suppose that $AA2$ holds. It implies that for any
$(A_{n})\in \mathcal{A}_{1}$, $\bar{Q}^{n}(A_{n})\nrightarrow 1$
holds. Taking $\mathbb{H}\equiv 1$ we obtain
\begin{equation*}
\tilde{v}(1)=\inf_{(A_{n})\in \mathcal{A}_{1}}\underset{n}{\underline{\lim }}%
\ \bar{Q}^{n}(A_{n})<1=v(1).
\end{equation*}
\end{rem}

\begin{rem}
\label{rem o zerowaniu ceny} If there is $SAA1$ or equivalently
$SAA2$, then for any $\mathbb{H}$ bounded, i.e. $H_{n}\leq K$ for
some constant $K>0$, we have $v_{\alpha }(\mathbb{H})=0$ for any
$\alpha \in \lbrack 0,1]$. Indeed,
by Theorem \ref{tw o arbitrazu asymtotycznym} there exists a sequence $(%
\tilde{A}_{n})$ s.t. $P^{n}(\tilde{A}_{n})\longrightarrow 1$ and $\bar{Q}%
^{n}(\tilde{A}_{n})\longrightarrow 0$. Then we have
\begin{equation*}
v_{\alpha }(\mathbb{H})\leq \tilde{v}(\mathbb{H})\leq \underset{n}{%
\underline{\lim }}\sup_{Q\in \mathcal{Q}^{n}}\mathbf{E}^{Q^{n}}[H_{n}\mathbf{%
1}_{\tilde{A}_{n}}]\leq \underset{n}{\underline{\lim }}\ K\ \bar{Q}^{n}(%
\tilde{A}_{n})=0.
\end{equation*}
\end{rem}

\noindent The next theorem provides some insight into the problem
of asymptotic pricing for complete models.

\begin{tw}
\label{tw o rownosci cen} Under the following assumptions:\newline
a) \ (NAA2) ,\newline b) \ the large market is complete, i.e.
$\mathcal{Q}^{n}=\{Q^{n}\}$ is a singleton for each $n$,\newline
c) \ $\mathbb{H}$ is bounded, i.e. $H_{n}\leq K$, for all $n$,
where $K$ is a positive constant,\newline we have
$v(\mathbb{H})=\tilde{v}(\mathbb{H})$.
\end{tw}
{\bf Proof :} First notice, that for any fixed
$(A_{n})\in\mathcal{A}_1$ by $NAA2$ we obtain
\begin{gather*}
P^{n}(A_{n})\longrightarrow 1\quad\Longleftrightarrow\quad
P^{n}(A^c_{n})\longrightarrow 0\quad\Longrightarrow\quad
Q^{n}(A^c_{n})\longrightarrow 0.
\end{gather*}
Now consider two sequences:
\begin{align*}
x_{n}&:=\mathbf{E}^{Q^n}[H_{n}] \\[2ex]
y_{n}&:=\mathbf{E}^{Q^n}[H_{n}\mathbf{1}_{A_{n}}].
\end{align*}
The following holds:
\begin{gather*}
x_{n}-y_{n}=\mathbf{E}^{Q^n}[H_{n}-H_{n}\mathbf{1}_{A_{n}}]=\mathbf{E}%
^{Q^n}[H_{n}\mathbf{1}_{A^c_{n}}] \leq K\cdot
Q^{n}(A^c_{n})\longrightarrow 0
\end{gather*}
and thus
\begin{gather*}
\underset{n}{\underline{\lim}} \
x_{n}=\underset{n}{\underline{\lim}} \ y_{n}.
\end{gather*}
Taking infimum over all $(A_n)\in\mathcal{A}_1$ we obtain the
required result.
\begin{gather*}
v(\mathbb{H})=\underset{n}{\underline{\lim}} \ x_{n}= \inf_{(A_{n})\in%
\mathcal{A}} \ \underset{n}{\underline{\lim}} \
y_{n}=\tilde{v}(\mathbb{H})
\end{gather*}
$\hfill\square$\newline

\begin{rem}
Assume that $NAA2$ holds. For incomplete market we can define the
analogous sequences as in Theorem \ref{tw o rownosci cen}:
\begin{align*}
x_{n}& :=\sup_{Q\in \mathcal{Q}^{n}}\mathbf{E}^{Q}[H_{n}] \\[2ex]
y_{n}& :=\sup_{Q\in
\mathcal{Q}^{n}}\mathbf{E}^{Q}[H_{n}\mathbf{1}_{A_{n}}]
\end{align*}
and for these sequences we obtain analogous inequality
\begin{equation*}
x_{n}-y_{n}\leq \sup_{Q\in \mathcal{Q}^{n}}\mathbf{E}^{Q}[H_{n}-H_{n}\mathbf{%
1}_{A_{n}}]=\sup_{Q\in \mathcal{Q}^{n}}\mathbf{E}^{Q}[H_{n}\mathbf{1}%
_{A_{n}^{c}}]\leq K\cdot \bar{Q}^{n}(A_{n}^{c}).
\end{equation*}
However, we do not know if the last term goes to $0$ as
$n\longrightarrow \infty $. We know that
$\underline{Q}^{n}(A_{n}^{c})\longrightarrow 0$ only and this is
insufficient to perform the above proof for incomplete markets.
\end{rem}

\section{The large Black-Scholes model}

\label{5} Let $W_{t}^{1},W_{t}^{2},...$ be a sequence of
independent
standard Brownian motions on a filtered probability space $(\Omega ,\mathcal{%
F}_{t},\mathcal{F},P),t\in \lbrack 0,T]$. We will consider a
stationary
market, where the $n$-th small market has its natural filtration i.e. $%
\mathcal{F}_{t}^{n}=\sigma ((W_{s}^{1},...,W_{s}^{n})_{s\in
\lbrack 0,t]})$ and $\mathcal{F}^{n}=\mathcal{F}_{T}^{n}$. The
$n$-th objective measure is an adequate restriction of $P$ i.e.
$P^{n}=P\mid _{\mathcal{F}^{n}}$ and the discounted price
processes are given by
\begin{equation*}
dS_{t}^{i}=S_{t}^{i}(b_{i}dt+\sigma _{i}dW_{t}^{i})\qquad
i=1,2,...,n,\quad t\in \lbrack 0,T]
\end{equation*}
where $b_{i}\in \mathbb{R},\sigma _{i}>0$ for $i=1,2,...,n$. Such
sequence forms a complete large market with martingale measures
given by densities
\begin{equation*}
\frac{dQ^{n}}{dP^{n}}=Z_{n}=e^{-(\mathbf{\theta }^{n},\mathbf{W}_{T}^{n})-%
\frac{1}{2}\parallel \mathbf{\theta }^{n}\parallel ^{2}T}
\end{equation*}
where $\mathbf{\theta }^{n}=(\frac{b_{1}}{\sigma _{1}},...,\frac{b_{n}}{%
\sigma _{n}})$ and $\mathbf{W}_{t}^{n}=(W_{t}^{1},...,W_{t}^{n})$.
Recall, that $\mathbf{W^{\ast
}}_{t}^{n}=\mathbf{W}_{t}^{n}+\mathbf{\theta }^{n}t$ is a Brownian
motion under $Q^{n}$. In this setting we show more indirect proofs
for the absence of asymptotic arbitrage using methods of
mathematical statistics for searching optimal non-randomized tests
(see Lemma \ref{lem NP}). The shortcoming of this approach is that
it works for deterministic coefficients only. In this section we
show also, that Theorem \ref{tw o rownosci cen} and Remark \ref{rem
o zerowaniu ceny} remain true for random variables satisfying some
integrability conditions, which are satisfied for widely used
derivatives.\newline

\noindent For this section use let us introduce a class of sequences $%
(\varepsilon _{n})$ which take values in the interval [0,1] and
converging to $0$. Such class will be denoted by $\mathcal{E}$.

\begin{tw}
\label{tw NAA1 BS} For $\varepsilon >0$ let $A_{\varepsilon }^{n}$
denote a solution of the problem
\begin{equation*}
A\in \mathcal{F}^{n}:
\begin{cases}
P^{n}(A)\longrightarrow \max  \\
Q^{n}(A)\leq \varepsilon .
\end{cases}
\end{equation*}
Then the following conditions are equivalent

\begin{enumerate}[1)]
\item  $NAA1$
\item  $(P^{n})\vartriangleleft ({Q}^{n})$
\item  For any sequence $(\varepsilon _{n})\in \mathcal{E}$,  $%
P^{n}(A_{\varepsilon _{n}}^{n})\longrightarrow 0$ holds.
\item
$\sum_{i=1}^{\infty }(\frac{b_{i}}{\sigma _{i}})^{2}<\infty $
\end{enumerate}
\end{tw}
{\bf Proof : } Equivalence of $(1)$ and $(2)$ is proved in ~\cite{KabKra2}%
. \newline
$(2)\Longrightarrow (3)$ Let $(\varepsilon _{n})$ be any element of $%
\mathcal{E}$. Then $Q^{n}(A_{\varepsilon _{n}}^{n})\leq \varepsilon
_{n}\longrightarrow 0$ and thus by $(2)$, $P^{n}(A_{\varepsilon
_{n}}^{n})\longrightarrow 0$ holds. \newline
$(3)\Longrightarrow (2)$ Let $A_{n}\in \mathcal{F}^{n}$ be s.t. $%
Q^{n}(A_{n})\longrightarrow 0$. Then $\varepsilon
_{n}:=Q^{n}(A_{n})$ belongs to class $\mathcal{E}$ and by $(3)$,
$P^{n}(A_{n})\leq P^{n}(A_{\varepsilon _{n}}^{n})\longrightarrow 0$
holds. \newline $(3)\Longleftrightarrow (4)$ Statistical methods
provide an explicit form of the set $A_{\varepsilon }^{n}$.
According to the Neyman-Pearson Lemma \ref {lem NP} it is of the
form $A_{\varepsilon }^{n}=\{\frac{dP^{n}}{dQ^{n}}\geq \gamma \}$,
where $\gamma $ is a constant s.t. $Q^{n}(A_{\varepsilon
}^{n})=\varepsilon $. This construction provides
\begin{align*}
A_{\varepsilon }^{n}=& \left\{ e^{(\mathbf{\theta }^{n},\mathbf{W}_{T}^{n})+%
\frac{1}{2}\parallel \mathbf{\theta }^{n}\parallel ^{2}T}\geq
\gamma
\right\} =\left\{ (\mathbf{\theta }^{n},\mathbf{W}_{T}^{n})\geq \ln \gamma -%
\frac{1}{2}\parallel \mathbf{\theta }^{n}\parallel ^{2}T\right\}  \\[2ex]
=& \left\{ (\mathbf{\theta }^{n},(\mathbf{W}_{T}^{\ast n}-\mathbf{\theta }%
T))\geq \ln \gamma -\frac{1}{2}\parallel \mathbf{\theta
}^{n}\parallel ^{2}T\right\} =\left\{ (\mathbf{\theta
}^{n},\mathbf{W}_{T}^{\ast n})\geq \ln \gamma
+\frac{1}{2}\parallel \mathbf{\theta }^{n}\parallel ^{2}T\right\}
.
\end{align*}
Solving the following equation:
\begin{equation*}
Q^{n}(A_{\varepsilon }^{n})=Q^{n}\bigg\{(\mathbf{\theta }^{n},\mathbf{W}%
_{T}^{\ast n})\geq \ln \gamma +\frac{1}{2}\parallel \mathbf{\theta }%
^{n}\parallel ^{2}T\bigg\}=1-\Phi \left( \frac{ln\gamma +\frac{1}{2}%
\parallel \mathbf{\theta }^{n}\parallel ^{2}T}{\parallel \mathbf{\theta }%
^{n}\parallel \sqrt{T}}\right) =\varepsilon
\end{equation*}
we obtain
\begin{equation*}
\gamma =e^{\parallel \mathbf{\theta }^{n}\parallel \sqrt{T}\Phi
^{-1}(1-\varepsilon )-\frac{1}{2}\parallel \mathbf{\theta
}^{n}\parallel ^{2}T}.
\end{equation*}
We calculate the value $P^{n}(A_{\varepsilon }^{n})$.
\begin{align*}
P^{n}(A_{\varepsilon }^{n})=& P^{n}\left( (\mathbf{\theta }^{n},\mathbf{W}%
_{T}^{n})\geq \ln \gamma -\frac{1}{2}\parallel \mathbf{\theta
}^{n}\parallel
^{2}T\right) =1-\Phi \left( \frac{ln\gamma -\frac{1}{2}\parallel \mathbf{%
\theta }^{n}\parallel ^{2}T}{\parallel \mathbf{\theta }^{n}\parallel \sqrt{T}%
}\right)  \\[2ex]
=& 1-\Phi \left( \frac{\parallel \mathbf{\theta }^{n}\parallel
\sqrt{T}\
\Phi ^{-1}(1-\varepsilon )-\parallel \mathbf{\theta }^{n}\parallel ^{2}T}{%
\parallel \mathbf{\theta }^{n}\parallel \sqrt{T}}\right) =1-\Phi \left( \Phi
^{-1}(1-\varepsilon )-\parallel \mathbf{\theta }^{n}\parallel \sqrt{T}%
\right)
\end{align*}
Now observe that if $\sum_{i=1}^{\infty }(\frac{b_{i}}{\sigma _{i}}%
)^{2}<\infty $ then for any $(\varepsilon _{n})\in \mathcal{E}$
\begin{equation*}
1-\Phi \left( \Phi ^{-1}(1-\varepsilon _{n})-\parallel \mathbf{\theta }%
^{n}\parallel \sqrt{T}\right) \longrightarrow 0.
\end{equation*}
If $\sum_{i=1}^{\infty }(\frac{b_{i}}{\sigma _{i}})^{2}=\infty $ then $%
\varepsilon _{n}:=1-\Phi (1+\parallel \theta ^{n}\parallel \sqrt{T}%
)\longrightarrow 0$ and
\begin{equation*}
1-\Phi \left( \Phi ^{-1}(1-\varepsilon _{n})-\parallel \mathbf{\theta }%
^{n}\parallel \sqrt{T}\right) =1-\Phi (1)\nrightarrow 0.
\end{equation*}
\hfill {$\square $}\newline The next two theorems provide
characterization of $NAA2$, $SAA1$ and $SAA2$. The proofs are
similar and therefore we sketch some parts of them only.

\begin{tw}
For $\varepsilon >0$ let $A_{\varepsilon }^{n}$ denote a solution of
the problem
\begin{equation*}
A\in \mathcal{F}^{n}:
\begin{cases}
Q^{n}(A)\longrightarrow \max  \\
P^{n}(A)\leq \varepsilon .
\end{cases}
\end{equation*}
Then the following conditions are equivalent
\begin{enumerate}[1)]
\item  $NAA2$
\item  $(Q^{n})\vartriangleleft ({P}^{n})$
\item  For any sequence $(\varepsilon _{n})\in \mathcal{E}$,  $%
Q^{n}(A_{\varepsilon _{n}}^{n})\longrightarrow 0$  holds. \item
$\sum_{i=1}^{\infty }(\frac{b_{i}}{\sigma _{i}})^{2}<\infty $
\end{enumerate}
\end{tw}
{\bf Proof :} $(3)\Longleftrightarrow (4)$ The set
$A^n_{\varepsilon}$ is of the form
\begin{gather*}
A^n_{\varepsilon}=\left\{\frac{dQ^n}{dP^n}\geq\gamma\right\}
\end{gather*}
where $\gamma$ is s.t. $P^n(A^n_{\varepsilon})=\varepsilon$. This
procedure yields
\begin{gather*}
A^n_{\varepsilon}=\left\{(\theta^n,\mathbf{W}^n_T)\leq\Phi\left(\frac{\ln%
\frac{1}{\gamma}- \frac{1}{2}\parallel\theta^n\parallel^2T}{%
\parallel\theta^n\parallel\sqrt{T}}\right)\right\} \\[2ex]
\gamma=e^{-\left[\Phi^{-1}(\varepsilon)\parallel\theta^n\parallel \sqrt{T}+%
\frac{1}{2}\parallel\theta^n\parallel^2T\right]}
\end{gather*}
and $Q_n(A^n_{\varepsilon})=\Phi\left(\Phi^{-1}(\varepsilon)+\parallel%
\theta^n\parallel \sqrt{T}\right)$.\newline
If $\sum_{i=1}^{\infty}(\frac{b_i}{\sigma_i})^2<\infty$ then for any $%
(\varepsilon_n)\in\mathcal{E}$
\begin{gather*}
\Phi\left(\Phi^{-1}(\varepsilon_n)+\parallel\theta^n\parallel \sqrt{T}%
\right) \longrightarrow 0.
\end{gather*}
If $\sum_{i=1}^{\infty}(\frac{b_i}{\sigma_i})^2=\infty$ then taking $%
\varepsilon_n:=\Phi(1-\parallel\theta^n\parallel
\sqrt{T})\longrightarrow 0$ we obtain
\begin{gather*}
\Phi\left(\Phi^{-1}(\varepsilon_n)+\parallel\theta^n\parallel \sqrt{T}%
\right)=\Phi(1)\nrightarrow 0.
\end{gather*}
\hfill{$\square$}\newline

\begin{tw}
For $\varepsilon >0$ let $A_{\varepsilon }^{n}$ denote a solution of
the problem
\begin{equation*}
A\in \mathcal{F}^{n}:
\begin{cases}
P^{n}(A)\longrightarrow \max  \\
Q^{n}(A)\leq \varepsilon .
\end{cases}
\end{equation*}
Then the following conditions are equivalent
\begin{enumerate}[1)]
\item $SAA1$ \item $SAA2$ \item $P^{n}\vartriangle {Q}^{n}$ \item
${Q}^{n}\vartriangle P^{n}$
\item  There exists a sequence $(\varepsilon _{n})\in \mathcal{E}\ s.t.$ $%
P^{n}(A_{\varepsilon _{n}}^{n})\longrightarrow 1$ \item
$\sum_{i=1}^{\infty }(\frac{b_{i}}{\sigma _{i}})^{2}=\infty $.
\end{enumerate}
\end{tw}
\noindent
Notice, that the conditions for the set $A_{\varepsilon
}^{n}$ are based on property $P^{n}\vartriangle {Q}^{n}$. One can
base the proof on the property
${Q}^{n}\vartriangle P^{n}$. This requires replacing measures $P^{n}$ and $%
Q^{n}$ in the conditions for $A_{\varepsilon }^{n}$. The first four
conditions are proved in ~\cite{KabKra2} and are included in the
formulation above for the clarity of exposition only. Equivalence of
$(3)$ and $(5) $ are easy to prove. \newline
{\bf Proof :} $(5)\Longleftrightarrow (6)$ We use the construction of $%
A_{\varepsilon }^{n}$ found in the proof of Th. \ref{tw NAA1 BS}
\begin{gather*}
A_{\varepsilon }^{n}=\left\{ (\mathbf{\theta
}^{n},\mathbf{W}_{T}^{\ast n})\geq \ln \gamma
+\frac{1}{2}\parallel \mathbf{\theta }^{n}\parallel
^{2}T\right\}  \\[2ex]
\gamma =e^{\parallel \mathbf{\theta }^{n}\parallel \sqrt{T}\Phi
^{-1}(1-\varepsilon )-\frac{1}{2}\parallel \mathbf{\theta
}^{n}\parallel
^{2}T} \\[2ex]
P^{n}(A_{\varepsilon }^{n})=1-\Phi \left( \Phi ^{-1}(1-\varepsilon
)-\parallel \mathbf{\theta }^{n}\parallel \sqrt{T}\right)
\end{gather*}
If $\sum_{i=1}^{\infty }(\frac{b_{i}}{\sigma _{i}})^{2}<\infty $
then for any $(\varepsilon _{n})\in \mathcal{E}$, $1-\Phi \left(
\Phi
^{-1}(1-\varepsilon _{n})-\parallel \mathbf{\theta }^{n}\parallel \sqrt{T}%
\right) \longrightarrow 0$ holds. If $\sum_{i=1}^{\infty }(\frac{b_{i}}{\sigma _{i}%
})^{2}=\infty $ then $\varepsilon _{n}:=1-\Phi
(\frac{1}{2}\parallel \theta ^{n}\parallel
\sqrt{T})\longrightarrow 0$ and $1-\Phi \left( \Phi
^{-1}(1-\varepsilon _{n})-\parallel \mathbf{\theta }^{n}\parallel \sqrt{T}%
\right) \longrightarrow 1$.\\ \phantom{}\quad \hfill {$\square $}

In the sequel we will characterize the weak price of $\mathbb{H}$
satisfying some integrability conditions. If $\mathbb{H}=H$, where
$H$ is one fixed random variable measurable with respect to
$\mathcal{F}^1$, then it is clear that $\mathbf{E}^{Q^n}[H]$ does
not depend on $n$ and thus indicates the strong price. This means
that the investor doesn't have any profits from the fact that the
market is getting large and that he can use greater and grater
number of strategies. It turns out that he can not make any profits
unless
he uses $1$-quantile hedging strategies. In this case, but if $%
\sum_{i=1}^{\infty}(\frac{b_i}{\sigma_i})^2=\infty$, the initial
endowment
can be reduced to 0, i.e. the weak price is equal to $0$. The condition $%
\sum_{i=1}^{\infty}(\frac{b_i}{\sigma_i})^2<\infty$ guaranties that
the investor is not able to make any profits at all, no matter what
strategies he uses, because then
$v(\mathbb{H})=\tilde{v}(\mathbb{H})$.

\begin{tw}
\label{tw dla BS} Let $\mathbb{H}$ be a contingent claim on a
large Black-Scholes market with constant coefficients. Then

\begin{enumerate}[1)]
\item  if $\sum_{i=1}^{\infty }(\frac{b_{i}}{\sigma _{i}})^{2}<\infty $ and $%
\overline{\underset{n}{\lim }}\ \mathbf{E}[H_{n}^{1+\delta
}]<\infty $ for some $\delta >0$ then
$\tilde{v}(\mathbb{H})=v(\mathbb{H})$.
\item  if $\sum_{i=1}^{\infty }(\frac{b_{i}}{\sigma _{i}})^{2}=\infty $ and $%
\overline{\underset{n}{\lim }}\ \mathbf{E}[H_{n}^{4+\delta
}]<\infty $ for some $\delta >0$ then $\tilde{v}(\mathbb{H})=0$.
\end{enumerate}
\end{tw}
{\bf Proof : } $(1)$ For any sequence $(A_{n})\in \mathcal{A}_{1}$
define
$x_{n}:=\mathbf{E}^{Q^{n}}[H_{n}]$, $y_{n}:=\mathbf{E}^{Q^{n}}[H_{n}\mathbf{1%
}_{A_{n}}]$. Let $p,q,p^{^{\prime }},q^{^{\prime }}>1$ be real
numbers and
s.t. $\frac{1}{p}+\frac{1}{q}=1$, $\frac{1}{p^{^{\prime }}}+\frac{1}{%
q^{^{\prime }}}=1$. Using H\"{o}lder inequality twice to the difference $%
x_{n}-y_{n}$ we obtain:
\begin{align*}
x_{n}-y_{n}=& \mathbf{E}^{Q^{n}}[H_{n}\mathbf{1}_{A_{n}^{c}}]=\mathbf{E}%
[Z_{n}H_{n}\mathbf{1}_{A_{n}^{c}}]\leq \Big(\mathbf{E}(Z_{n}H_{n})^{p}\Big)^{%
\frac{1}{p}}\Big(P(A_{n}^{c})\Big)^{\frac{1}{q}} \\[2ex]
\leq & \bigg(\Big(\mathbf{E}(Z_{n}^{pp^{^{\prime }}})\Big)^{\frac{1}{%
p^{\prime }}}\Big(\mathbf{E}(H_{n}^{pq^{^{\prime }}})\Big)^{\frac{1}{%
q^{\prime }}}\bigg)^{\frac{1}{p}}\Big(P(A_{n}^{c})\Big)^{\frac{1}{q}} \\[2ex]
=& \Big(\mathbf{E}(Z_{n}^{pp^{^{\prime }}})\Big)^{\frac{1}{pp^{\prime }}}%
\Big(\mathbf{E}(H_{n}^{pq^{^{\prime }}})\Big)^{\frac{1}{pq^{\prime }}}\Big(%
P(A_{n}^{c})\Big)^{\frac{1}{q}}.
\end{align*}
Straightforward calculations yields
\begin{gather}\label{wzor na w.o.}
\Big(\mathbf{E}(Z_{n}^{pp^{^{\prime }}})\Big)^{\frac{1}{pp^{\prime }}}=e^{%
\frac{1}{2}\parallel \theta ^{n}\parallel ^{2}T(pp^{^{\prime
}}-1)}
\end{gather}
and thus $NAA2$ guaranties that $\lim_{n\rightarrow \infty }\Big(\mathbf{E}%
(Z_{n}^{pp^{^{\prime }}})\Big)^{\frac{1}{pp^{\prime }}}<\infty $. Taking $%
p,p^{^{\prime }}$ s.t. $pq^{^{\prime }}=1+\delta $ and using fact that $%
\lim_{n\rightarrow \infty }\Big(P(A_{n}^{c})\Big)^{\frac{1}{q}}=0$
we
conclude that $\lim_{n\rightarrow \infty }(x_{n}-y_{n})=0$. Thus $\underline{%
\lim }\ x_{n}=\underline{\lim }\ y_{n}$ and taking infimum over all $%
(A_{n})\in \mathcal{A}_{1}$ we get $\tilde{v}(\mathbb{H})=v(\mathbb{H})$.%
\newline
\newline
\noindent $(2)$ For any $(A_{n})\in \mathcal{A}_{1}$,
$p,p^{^{\prime }}>1$
and $q,q^{^{\prime }}$ s.t. $\frac{1}{p}+\frac{1}{q}=1$, $\frac{1}{%
p^{^{\prime }}}+\frac{1}{q^{^{\prime }}}=1$ using H\"{o}lder
inequalities we obtain:
\begin{align*}
\mathbf{E}^{Q^{n}}[H_{n}\mathbf{1}_{A_{n}}]& \leq \Big(\mathbf{E}%
^{Q^{n}}(H_{n}^{p})\Big)^{\frac{1}{p}}\Big(Q^{n}(A_{n})\Big)^{\frac{1}{q}}=%
\Big(\mathbf{E}(Z_{n}H_{n}^{p})\Big)^{\frac{1}{p}}\Big(Q^{n}(A_{n})\Big)^{%
\frac{1}{q}} \\[2ex]
& \leq \bigg(\Big(\mathbf{E}Z_{n}^{p^{^{\prime }}}\Big)^{\frac{1}{%
p^{^{\prime }}}}\Big(\mathbf{E}H_{n}^{pq^{^{\prime }}}\Big)^{\frac{1}{%
q^{^{\prime }}}}\bigg)^{\frac{1}{p}}\Big(Q^{n}(A_{n})\Big)^{\frac{1}{q}}=%
\Big(\mathbf{E}Z_{n}^{p^{^{\prime }}}\Big)^{\frac{1}{pp^{^{\prime }}}}\Big(%
\mathbf{E}H_{n}^{pq^{^{\prime }}}\Big)^{\frac{1}{pq^{^{\prime }}}}\Big(%
Q^{n}(A_{n})\Big)^{\frac{1}{q}}
\end{align*}
Now, similarly to the previously used methods let us solve an
auxiliary problem of finding set $A_{\varepsilon }^{n}$ s.t.
\begin{equation*}
A\in \mathcal{F}^{n}:
\begin{cases}
Q^{n}(A)\longrightarrow \min  \\
P^{n}(A)\geq 1-\varepsilon .
\end{cases}
\end{equation*}
Analogous calculations provide:
\begin{gather*}
A_{\varepsilon }^{n}=\left\{ \frac{dQ^{n}}{dP^{n}}\leq \gamma
\right\} =\left\{ (\theta ^{n},\mathbf{W}^{n})\geq -\ln \gamma
\frac{1}{2}\parallel
\theta ^{n}\parallel ^{2}T\right\}  \\[2ex]
\gamma =e^{-[\Phi ^{-1}(\varepsilon )\parallel \theta ^{n}\parallel \sqrt{T}+%
\frac{1}{2}\parallel \theta ^{n}\parallel ^{2}T]} \\[2ex]
Q^{n}(A_{\varepsilon }^{n})=\Phi \Big(-\Phi ^{-1}(\varepsilon
)-\parallel \theta ^{n}\parallel \sqrt{T}\Big).
\end{gather*}
Taking $p=2+\frac{1}{2}\delta ,p^{^{\prime }}=2,\varepsilon _{n}=\Phi \Big(%
-\ln (\parallel \theta ^{n}\parallel \sqrt{T})\Big)$ (AA2 guaranties that $%
\varepsilon _{n}\rightarrow 0$) we get
\begin{equation*}
\overline{\lim }\ \mathbf{E}[H_{n}^{pq^{^{\prime
}}}]=\overline{\lim }\ \mathbf{E}[H_{n}^{4+\delta }]<\infty
\end{equation*}
and
\begin{align}
\bigg(\Big(\mathbf{E}Z_{n}^{p^{^{\prime }}}\Big)^{\frac{1}{pp^{^{\prime }}}}%
\Big(Q^{n}(A_{\varepsilon _{n}}^{n})\Big)^{\frac{1}{q}}\bigg)^{q}& =e^{\frac{%
1}{2}\frac{p^{^{\prime }}-1}{p-1}\parallel \theta ^{n}\parallel ^{2}T}\Phi %
\Big(-\Phi ^{-1}(\varepsilon _{n})-\parallel \theta ^{n}\parallel \sqrt{T}%
\Big)  \notag  \label{wz na wykladniki H} \\[2ex]
& =e^{\frac{1}{2+\delta }\parallel \theta ^{n}\parallel ^{2}T}\Phi
\Big(\ln (\parallel \theta ^{n}\parallel \sqrt{T})-\parallel
\theta ^{n}\parallel \sqrt{T}\Big)
\end{align}
Replacing $\parallel \theta ^{n}\parallel \sqrt{T}$ by $x$ for the
sake of convenience, we calculate the following limit using
d'Hospital formula.
\begin{align*}
\lim_{x\rightarrow \infty }e^{\frac{1}{2+\delta }x^{2}}\Phi (\ln
x-x)& =\lim_{x\rightarrow \infty }\frac{\frac{1}{\sqrt{2\pi
}}e^{-\frac{1}{2}(\ln
x-x)^{2}}(\frac{1}{x}-1)}{e^{-\frac{1}{2+\delta }x^{2}}(-\frac{1}{2+\delta }%
)2x} \\[2ex]
& =\lim_{x\rightarrow \infty }-\frac{2+\delta }{2\sqrt{2\pi }}\left[ \frac{%
e^{x^{2}(\frac{1}{2+\delta }-\frac{1}{2})-\frac{1}{2}\ln ^{2}x+x\ln x}}{x^{2}%
}-\frac{e^{x^{2}(\frac{1}{2+\delta }-\frac{1}{2})-\frac{1}{2}\ln
^{2}x+x\ln x}}{x}\right] =0
\end{align*}
The limit is equal to $0$ since: $\lim x^{2}(\frac{1}{2+\delta
}-\frac{1}{2})-\frac{1}{2}\ln ^{2}x+x\ln x=-\infty $.\\
Summarizing, we have shown that
$\lim_{n\rightarrow \infty }\Big(\mathbf{E}Z_{n}^{p^{^{\prime }}}\Big)^{%
\frac{1}{pp^{^{\prime }}}}\Big(\mathbf{E}H^{pq^{^{\prime }}}\Big)^{\frac{1}{%
pq^{^{\prime }}}}\Big(Q^{n}(A_{\varepsilon
_{n}}^{n})\Big)^{\frac{1}{q}}=0$ \ for the adequate parameters and
thus $\tilde{v}(\mathbb{H})=0$. \phantom{i} \hfill {$\square
$}\newline

\begin{rem}
The integrability conditions imposed on $\mathbb{H}$ in the second
item of Theorem \ref{tw dla BS} can be a little bit weakened. It
follows from \ref {wz na wykladniki H} that we have to find
parameters $p,p^{^{\prime }}>1$ s.t. $\frac{1}{2}\frac{p^{^{\prime
}}-1}{p-1}=\frac{1}{2+\delta }$. We can impose additional
requirement: $pq^{^{\prime }}\rightarrow \min $. Then it can be
checked, that the solution is: $p=1+\sqrt{\frac{%
2+\delta }{2}}$, $p^{^{\prime }}=\frac{2+2\sqrt{\frac{2+\delta
}{2}}+\delta
}{2+\delta }$ and $pq^{^{\prime }}=\frac{\sqrt{2}}{2}+\frac{1}{2}(4+\delta )+%
\frac{\sqrt{2}}{2}\sqrt{2+\delta }$. If $\delta \rightarrow 0$ then $%
pq^{^{\prime }}$ is arbitrarily close to $\frac{\sqrt{2}}{2}+2+1<4$.
Thus, we can assume that
\begin{equation*}
\overline{\underset{n}{\lim }}\ \mathbf{E}[H_{n}^{\frac{\sqrt{2}}{2}+\frac{1%
}{2}(4+\delta )+\frac{\sqrt{2}}{2}\sqrt{2+\delta }}]<\infty \ \text{for some}%
\ \delta >0.
\end{equation*}
\end{rem}

\noindent The next theorem provides a more precise characterization of the $%
\alpha $-quantile price. But first let us impose a regularity
assumption on the random variables $H_{n}Z_{n}$.

\begin{ass}
\label{ass o dystrybuancie} The random variable $H_{n}Z_{n}$ has a
continuous distribution function with respect to the measure
$P^{n}$.
\end{ass}
By $q_n(\alpha)$ we denote the $\alpha$-quantile of $H_nZ_n$, i.e.
$q_n(\alpha)=\{\inf x : P^n(H_nZ_z\leq x)\geq\alpha\}$.
\\
\\
\noindent Denote by $\mathcal{B}_{\alpha}$ a set of sequences
satisfying
\begin{equation*}
\underset{n\longrightarrow
\infty}{\underline{\lim}}\beta_n\geq\alpha.
\end{equation*}

\begin{tw}
\label{tw o cenie kwantylowej dla BS} Let $\mathbb{H}$ be a
contingent claim on a large Black-Scholes model with constant
coefficients.

\begin{enumerate}[1)]
\item  Under assumption \ref{ass o dystrybuancie} the $\alpha
$-quantile price is given by the formula
\begin{equation*}
v_{\alpha }(\mathbb{H})=\inf_{(\beta _{n})\in \mathcal{B}_{\alpha }}%
\underset{n\rightarrow \infty }{\underline{\lim }}\mathbf{E}\Big[H_{n}Z_{n}%
\mathbf{1}_{\{H_{n}Z_{n}\leq q_{n}(\beta _{n})\}}\Big].
\end{equation*}

\item  Let assumption \ref{ass o dystrybuancie} be satisfied. If $\overline{%
\underset{n}{\lim }}\ \mathbf{E}[H_{n}^{1+\delta }]<\infty $ for some $%
\delta >0$ and $\sum_{i=1}^{\infty }(\frac{b_{i}}{\sigma
_{i}})^{2}<\infty $ then
\begin{equation*}
v_{\alpha }(\mathbb{H})=\underset{n\rightarrow \infty }{\underline{\lim }}%
\mathbf{E}\Big[H_{n}Z_{n}\mathbf{1}_{\{H_{n}Z_{n}\leq q_{n}(\alpha
)\}}\Big].
\end{equation*}
Moreover, $v_{\alpha }(\mathbb{H})$ is a Lipschitz, increasing
function of $\alpha $ taking values in the interval
$[0,v(\mathbb{H})]$.\newline

\item  If $\overline{\underset{n}{\lim }}\
\mathbf{E}[H_{n}^{4+\delta
}]<\infty $ for some $\delta >0$ and $\sum_{i=1}^{\infty }(\frac{b_{i}}{%
\sigma _{i}})^{2}=\infty $ then $v_{\alpha }(\mathbb{H})=0$ for
each $\alpha \in \lbrack 0,1]$.
\end{enumerate}
\end{tw}
{\bf Proof:} (1) By Theorem \ref{tw o cenie kwantylowej} the $\alpha $%
-quantile price is given by the formula:
\begin{equation*}
v_{\alpha }(\mathbb{H})=\inf_{(A_{n})\in \mathcal{A}_{\alpha }}\ \underset{n%
}{\underline{\lim }}\ \sup_{Q\in \mathcal{Q}^{n}}\mathbf{E}^{Q}[H_{n}\mathbf{%
1}_{A_{n}}].
\end{equation*}
Let us consider any $(A_{n})\in \mathcal{A}_{\alpha }$ and define
$\beta _{n}:=P^{n}(A_{n})$. Denote by $\tilde{A}_{n}$ a solution
of the following problem:
\begin{equation*}
\tilde{A}_{n}:
\begin{cases}
\mathbf{E}^{Q^{n}}[H_{n}\mathbf{1}_{A_{n}}]\longrightarrow \min  \\
P^{n}(A_{n})\geq \beta _{n}.
\end{cases}
\end{equation*}
If we introduce measure $\tilde{Q}^{n}$ by the density $\frac{d\tilde{Q}^{n}%
}{dQ^{n}}:=\frac{H_{n}}{\mathbf{E}^{Q^{n}}[H_{n}]}$, then the
above problem can be written in the equivalent form:
\begin{equation*}
\tilde{A}_{n}:
\begin{cases}
\tilde{Q}^{n}(A_{n})\longrightarrow \min  \\
P^{n}(A_{n})\geq \beta _{n}.
\end{cases}
\end{equation*}
Therefore by Lemma \ref{lem NP} we conclude that $\tilde{A}_{n}$ is
of the
form: $\{H_{n}Z_{n}\leq \gamma \}$, where $\gamma $ is a constant s.t. $%
P^{n}(H_{n}Z_{n}\leq \gamma )=\beta _{n}$. By Assumption \ref{ass o
dystrybuancie} we know that there exists such $\gamma $ and it is equal to $%
q_{n}(\beta _{n})$. Thus $\tilde{A}_{n}=\{H_{n}Z_{n}\leq
q_{n}(\beta _{n})\}$ and
\begin{equation*}
\mathbf{E}^{Q^{n}}[H_{n}\mathbf{1}_{A_{n}}]\geq \mathbf{E}^{Q^{n}}[H_{n}%
\mathbf{1}_{\{H_{n}Z_{n}\leq q_{n}(\beta _{n})\}}].
\end{equation*}
Letting $n\rightarrow \infty $ and taking infimum over all
$(A_{n})\in \mathcal{A}_{\alpha }$ we obtain:
\begin{gather}\label{wz nierownosc}
v_{\alpha }(\mathbb{H})\geq \inf_{(\beta _{n})\in
\mathcal{B}_{\alpha }} \underset{n\rightarrow \infty
}{\underline{\lim }}\mathbf{E}\Big[H_{n}Z_{n}
\mathbf{1}_{\{H_{n}Z_{n}\leq q_{n}(\beta _{n})\}}\Big].
\end{gather}
However, $P^{n}(H_{n}Z_{n}\leq q_{n}(\beta _{n}))=\beta _{n}$, so $%
\{H_{n}Z_{n}\leq q_{n}(\beta _{n})\}\in \mathcal{A}_{\alpha }$ and
this implies equality in \ref{wz nierownosc}.\newline

\noindent (2) Let $\alpha ,\beta \in \lbrack 0,1]$ be two real numbers s.t. $%
\beta <\alpha $. For $p,q,p^{^{\prime }},q^{^{\prime }}>1$ s.t. $\frac{1}{p}+%
\frac{1}{q}=1,\frac{1}{p^{^{\prime }}}+\frac{1}{q^{^{\prime }}}=1$
we have the following inequality:
\begin{align*}
\mathbf{E}[H_{n}Z_{n}\mathbf{1}_{\{H_{n}Z_{n}\leq q_{n}(\alpha )\}}]-\mathbf{%
E}[H_{n}Z_{n}\mathbf{1}_{\{H_{n}Z_{n}\leq q_{n}(\beta )\}}]& =\mathbf{E}%
[H_{n}Z_{n}\mathbf{1}_{\{q_{n}(\beta )\leq H_{n}Z_{n}\leq
q_{n}(\alpha )\}}]
\\[2ex]
\leq \Big(\mathbf{E}(Z_{n}^{pp^{^{\prime }}})\Big)^{\frac{1}{pp^{\prime }}}%
\Big(\mathbf{E}(H_{n}^{pq^{^{\prime }}})\Big)^{\frac{1}{pq^{\prime }}}\Big(P(%
{q_{n}(\beta )\leq H_{n}Z_{n}\leq q_{n}(\alpha )})\Big)^{\frac{1}{q}}& =\Big(%
\mathbf{E}(Z_{n}^{pp^{^{\prime }}})\Big)^{\frac{1}{pp^{\prime }}}\Big(%
\mathbf{E}(H_{n}^{pq^{^{\prime }}})\Big)^{\frac{1}{pq^{\prime
}}}(\alpha -\beta )
\end{align*}
However, by \ref{wzor na w.o.} we have
$\Big(\mathbf{E}(Z_{n}^{pp^{^{\prime
}}})\Big)^{\frac{1}{pp^{\prime }}}\leq \lim_{n\rightarrow \infty }\Big(%
\mathbf{E}(Z_{n}^{pp^{^{\prime }}})\Big)^{\frac{1}{pp^{\prime
}}}<\infty $.
Taking $p,q^{^{\prime }}$ s.t. $pq^{^{\prime }}=1+\delta $ and denoting $%
K_{1}:=\lim_{n\rightarrow \infty }\Big(\mathbf{E}(Z_{n}^{pp^{^{\prime }}})%
\Big)^{\frac{1}{pp^{\prime }}}$ and $K_{2}:=\Big(\mathbf{E}%
(H_{n}^{pq^{^{\prime }}})\Big)^{\frac{1}{pq^{\prime }}}$, we
obtain
\begin{equation*}
\mathbf{E}[H_{n}Z_{n}\mathbf{1}_{\{H_{n}Z_{n}\leq q_{n}(\alpha )\}}]-\mathbf{%
E}[H_{n}Z_{n}\mathbf{1}_{\{H_{n}Z_{n}\leq q_{n}(\beta )\}}]\leq
K_{1}K_{2}(\alpha -\beta ).
\end{equation*}
and interchanging the role of $\ \alpha $ and $\beta $\ we obtain
\begin{gather}\label
{wz oszacowanie2}
\mid \mathbf{E}[H_{n}Z_{n}\mathbf{1}_{\{H_{n}Z_{n}\leq q_{n}(\alpha )\}}]-%
\mathbf{E}[H_{n}Z_{n}\mathbf{1}_{\{H_{n}Z_{n}\leq q_{n}(\beta
)\}}]\mid \leq K_{1}K_{2}\mid \alpha -\beta \mid .
\end{gather}
\noindent Now consider $(\beta _{n})\in \mathcal{B}_{\alpha }$. If $%
\underset{n\longrightarrow \infty }{\underline{\lim }}\beta
_{n}>\alpha $ then
$\mathbf{E}[H_{n}Z_{n}\mathbf{1}_{\{H_{n}Z_{n}\leq q_{n}(\beta
_{n})\}}]>\mathbf{E}[H_{n}Z_{n}\mathbf{1}_{\{H_{n}Z_{n}\leq
q_{n}(\alpha
)\}}]$ and thus $\underline{\lim }\ \mathbf{E}[H_{n}Z_{n}\mathbf{1}%
_{\{H_{n}Z_{n}\leq q_{n}(\beta _{n})\}}]\geq \underline{\lim }\ \mathbf{E}%
[H_{n}Z_{n}\mathbf{1}_{\{H_{n}Z_{n}\leq q_{n}(\alpha )\}}]$. If $\underset{%
n\longrightarrow \infty }{\underline{\lim }}\beta _{n}=\alpha $
then by \ref
{wz oszacowanie2} we have $\mid \mathbf{E}[H_{n}Z_{n}\mathbf{1}%
_{\{H_{n}Z_{n}\leq q_{n}(\alpha )\}}]-\mathbf{E}[H_{n}Z_{n}\mathbf{1}%
_{\{H_{n}Z_{n}\leq q_{n}(\beta _{n})\}}]\mid \leq K_{1}K_{2}\mid
\alpha
-\beta _{n}\mid $ and letting $n\rightarrow \infty $ we obtain $%
\underline{\lim }\
\mathbf{E}[H_{n}Z_{n}\mathbf{1}_{\{H_{n}Z_{n}\leq
q_{n}(\beta _{n})\}}]=\underline{\lim }\ \mathbf{E}[H_{n}Z_{n}\mathbf{1}%
_{\{H_{n}Z_{n}\leq q_{n}(\alpha )\}}]$. The conclusion from these
two cases
is that $v_{\alpha }(\mathbb{H})\geq \underline{\lim }\ \mathbf{E}[H_{n}Z_{n}%
\mathbf{1}_{\{H_{n}Z_{n}\leq q_{n}(\alpha )\}}]$. However,
$\{H_{n}Z_{n}\leq q_{n}(\alpha )\}\in \mathcal{A}_{\alpha }$ and
therefore
\begin{gather}\label{wz na alfa cene w BS}
v_{\alpha }(\mathbb{H})=\underset{n\rightarrow \infty }{\underline{\lim }}%
\mathbf{E}\Big[H_{n}Z_{n}\mathbf{1}_{\{H_{n}Z_{n}\leq q_{n}(\alpha
)\}}\Big].
\end{gather}
Letting $n\rightarrow \infty $ in \ref{wz oszacowanie2} and using
\ref{wz na alfa cene w BS} we obtain:
\begin{equation*}
\mid v_{\alpha }(\mathbb{H})-v_{\beta }(\mathbb{H})\mid \leq
K_{1}K_{2}\mid \alpha -\beta \mid ,
\end{equation*}
which proves that $v_{\alpha }(\mathbb{H})$ is Lipschitz. It is
clear by \ref {wz na alfa cene w BS} that $v_{\alpha }(\mathbb{H})$
is increasing and that $v_{0}(\mathbb{H})=0$. By Theorem \ref{tw dla
BS} $v_{1}(\mathbb{H})=v(\mathbb{H})$.
\newline
(3) It is an immediate consequence of Theorem \ref{tw dla BS} (2), since $%
v_{\alpha }(\mathbb{H})\leq \tilde{v}(\mathbb{H})$. \hfill
$\square $

\begin{rem}
Consider the prices of a call option, i.e. $\mathbb{H}\equiv
(S_{T}^{1}-K)^{+}$. The distribution of $(S_{T}^{1}-K)^{+}Z_{n}$ is
discontinuous in $0$. Let $\alpha _{0}:=P^{n}((S_{T}^{1}-K)^{+}=0)$.
It is clear, that for $\alpha \leq \alpha _{0}$, $v_{\alpha
}(\mathbb{H})=0$ holds. On the interval $(0,\infty )$ the
distribution function is continuous, thus for $\alpha >\alpha _{0}$
Theorem \ref{tw o cenie kwantylowej dla BS} can be applied.
\end{rem}

\noindent {\bf Conclusion}\\
 In this paper we have introduced and
characterized two types of asymptotic prices. They are based on
different treating of hedging risk which disappears in infinity.
Relations between them strictly depend on the asymptotic arbitrage
on the market. In case of the large Black-Scholes model with
constant coefficients it was possible to find more indirect formula
for the $\alpha$-quantile price and state some properties of it. On
this market there are two situations possible: \\
1) there is no asymptotic arbitrage of any kind - then the strong
and
the weak price are equal\\
2) there is asymptotic arbitrage of all kinds - then the weak price
is equal to zero, while the strong is not.\\

\noindent {\bf Acknowledgement} This paper is a part of the
author's PhD thesis. The author would like to express his thanks
to Professor Łukasz Stettner for help and support while writing
this paper.

\thispagestyle{empty}


\begin{thebibliography}{99}
\bibitem{DelSch}    Delbaen F., Schachermayer W. \thinspace\ \emph{The
fundamental Theorem of Asset Pricing for unbounded stochastic
processes}, \ Mathematische Annalen 312, 215-250, (1998)

\bibitem{FolKab}    F\"{o}llmer H., Kabanov Yu.M. \thinspace\ \emph{Optional
decomposition and Lagrange multipliers} Finance Stoch.- 2, 69-81
(1998)

\bibitem{FolKra}    F\"{o}llmer H., Kramkov D.O. \thinspace\ \emph{Optional
decompositions under constraints} Probab. Theory Relat. Fields
109, 1-25 (1997)

\bibitem{FolLeu}    F\"{o}llmer H., Leukert P.\thinspace\ \emph{Quantile
Hedging}, \  Finance and Stochastics 3, 251-273, (1999)

\bibitem{JacShi1}   Jacod J., Shiryaev A.N.\thinspace\ \emph{Limit Theorems
for Stochastic Processes}, Berlin Heidelberg New York : Springer
(1987)

\bibitem{JacShi2}   Jacod J., Shiryaev A.N. \thinspace\ \emph{Local
martingales and the fundamental asset pricing theorems in the
discrete-time case}, \  Finance and Stochastics 2, 259-273, (1998)

\bibitem{KabKra1}   Kabanov Yu.M., Kramkov D.O. \thinspace\ \emph{Large
financial markets: asymptotic arbitrage and contiguity}, Prob.
Theory Appl. 39(1), 182-187 (1994)

\bibitem{KabKra2}   Kabanov Yu.M., Kramkov D.O. \thinspace\ \emph{Asymptotic
arbitrage in large financial markets}, Finance Stoch. 2(2),
143-172 (1998)

\bibitem{Klein}     Klein I., \emph{A fundamental theorem of asset pricing for
large financial markets}, Mathematical Finance 10(4), 443-458
(2000)

\bibitem{KleScha}   Klein I., Schachermayer W. \thinspace\ \emph{Asymptotic
arbitrage in non-complete large financial markets}, Probab. Theory
Appl. 41(4), 780-788 (1996)

\bibitem{Kra}       Kramkov D.O. \thinspace\ \emph{Optional decomposition of
supermartingales and hedging contingent claims in incomplete security markets%
}, Probab. Theory Relat. Fields 105, 459-479 (1996)

\bibitem{Ras}   R\'{a}sonyi M. \thinspace\ \emph{On certain problems of
arbitrage theory in discrete time financial market models}, PhD
thesis (2002)
\end{thebibliography}
\end{document}